# Research of time discrimination circuits for PMT signal readout over large dynamic range in LHAASO WCDA

**C. Ma,**[a,b] **L. Zhao**[a,b,1], **R. Dong**[a,b], **Z. Jiang**[a,b], **S. Chu**[a,b], **X. Gao**[a,b], **S. Liu**[a,b], **Q. An**[a,b]

[a] *State Key Laboratory of Particle Detection and Electronics, University of Science and Technology of China,*
*Jinzhai Road, Hefei, 230026, China*

[b] *Department of Modern Physics, University of Science and Technology of China,*
*Jinzhai Road, Hefei, 230026, China*

*E-mail:* zlei@ustc.edu.cn

ABSTRACT: In the readout electronics of the Water Cerenkov Detector Array (WCDA) in the Large High Altitude Air Shower Observatory (LHAASO), both high-resolution charge and time measurement are required over a dynamic range from 1 photoelectron (P.E.) to 4000 P.E. for the PMT signal readout. In this paper, we present our work on the design of time discrimination circuits in LHAASO WCDA, especially on improvement to reduce the circuit dead time. Several approaches were studied through analysis and simulations, and actual circuits were designed and tested in the laboratory to evaluate the performance. Test results indicate that a time resolution better than 500 ps RMS is achieved in the whole large dynamic range, and the circuit dead time is successfully reduced to less than 200 ns.

---

[1] Corresponding author.

# Contents

# 1. Introduction

The Large High Altitude Air Shower Observatory (LHAASO) project is proposed to be built as a new generation complexity at a high altitude of more than 4,000 meters in China, with the aim of detecting the high energy extensive air showers (EAS). [1, 2] The Water Cerenkov Detector Array (WCDA) is an important component in LHAASO, and 3600 photomultiplier tubes (PMTs) are scattered in four individual same-sized square water ponds of 150 m ×150 m to detect Cerenkov light emitted by the EAS in the water. [3] Each front-end electronics (FEE) module above the water ponds recieves the signals from adjacent 9 PMTs through a 30-meter cable, and both high-resolution time and charge measurement are required over a large dynamic range from 1 photoelectron (1 P.E.) to 4000 P.E. The schematic drawing of the WCDA layout is shown in figure Figure 1. For the reconstruction of the physics event, the time measurement resolution should be better than 0.5 ns RMS, and the charge measurement resolution should be better than 30% @ 1 P.E. and 3% @4000 P.E. [4]

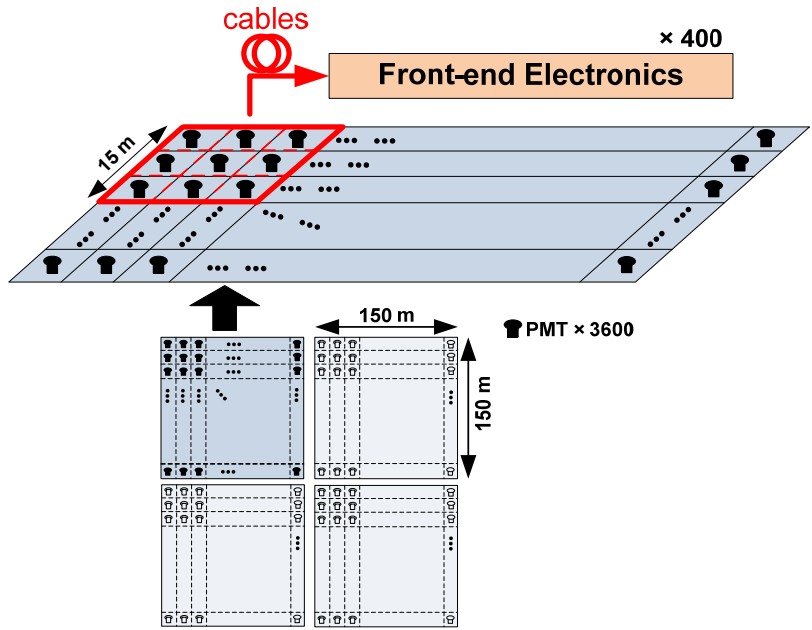

**Figure 1** Schematic drawing of the WCDA layout.

According to the application requirements, a prototype FEE shown in figure Figure 2 has been designed. Because the large dynamic range exceeds the capability of one single charge measurement electronics channel, two independent charge measurement channels are employed for one PMT (1 P.E. ~ 133 P.E. for the anode; 30 P.E. ~ 4000 P.E. for the 8$^{th}$ dynode), and the anode output signal is used to achieve the time measurement over the whole dynamic range.

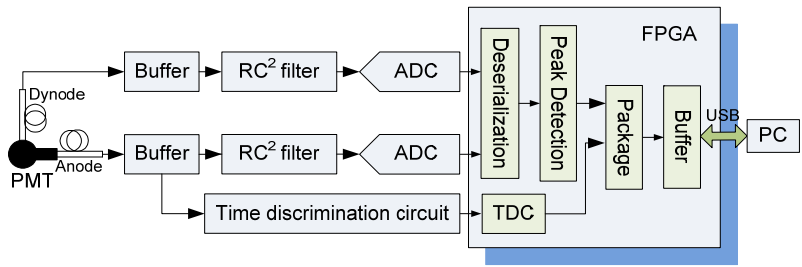

**Figure 2.** Block diagram of the prototype FEE module.

The charge measurement circuit is based on the analog shaping and digital peak detection method. The analogue shaper consists of a two-stage low-pass filter (RC$^2$ filter), and the time constant is set to 40 ns based on the analysis of the signal-to-noise ratio (SNR), peak error and other parameters obtained from the circuit simulations. Ref [5] shows more details about this. The quasi-Gaussian output signal of the shaper is fed to an Analog-to-Digital Converter (ADC), and the digitalized data stream is fed to the peak detection logic implemented in the FPGA devices (Cyclone III, from Altera Inc.) to acquire the charge information.

For the time measurement, a leading edge discriminator is combined with a FPGA-TDC (bin size: ~ 333 ps) based on the multi-phase clock interpolation technique [6] to reduce the system complexity. Finally, the time and charge measurement results are packed into a buffer and then transmitted to PC through the USB interface.

Efforts have been devoted to optimize the performance of the time discrimination circuit, which will be discussed in detail below. This paper is organized as follows: Section 2 introduces several approaches to increase the time resolution and dead time performance, including the circuit analysis and simulation; Section 3 presents a series of electronics test results to evaluate the circuit performance; finally, in Section 4, we conclude the paper and summarize what has been achieved.

## 2. Design of the time discrimination circuit

### 2.1 Previous time discrimination circuit

According to the above mentioned measurement requirements, the main design considerations of the time discrimination circuit include:

(1) Large dynamic range measurement. The time discrimination circuit receives the current signals from the anode over the whole dynamic range. In order to simplify the engineering design, signals from PMT are directly supplied to the readout circuit through a long cable without a pre-amplifier, and the voltage amplitude of the signal on the 50 Ω is as low as 3 mV for 1 P.E. while 5 V for 4000 P.E. due to the output saturation of the anode. Therefore, we should consider the dynamic response of the circuit for both the small and large input signals.

(2) High-precision impedance matching. Considering the large measurement dynamic range and the cable between the PMT and the time discrimination circuit, high-precision impedance matching (50 Ω) should be achieved to reduce the interference introduced by the reflection of the large pulse signals. A high-precision resistor of 50 Ω ($R_t$, shown in figure Figure 3) is employed to achieve the terminal impedance matching. As for the large input signals, the on-resistance of the protection diodes ($D_1$) would apparently decrease the equivalent impedance at the non-inverting input of the following operational amplifier ($A_1$, AD8000 from ADI Inc.). In this condition, the input impedance of the circuit mainly depends on ($R_1//R_t$). Therefore, a high-value resistor ($R_1$) is selected to guarantee the good impedance matching precision over the whole dynamic range.

(3) High time resolution measurement of the small signals. The amplitude of the 1 P.E. signal from the anode is very small. Therefore, the small signals should be amplified with a gain high enough to guarantee sufficient value of signal slew rate at the input of the discriminator. To meet this requirement, another high-speed amplifier ($A_2$, AD8000) is employed. On the other hand, the input offset voltage of AD8000 [7] would be also amplified, which would introduce a large DC bias at the input of the discriminator. Thus the amplified signals are AC coupled ($C_1$ and $R_2$, 1 nF × 50 Ω) to filter out the baseline drift and fluctuation, and then fed to a high-speed discriminator (MAX9601, from Maxim Inc.) with a threshold of 1/3 P.E.

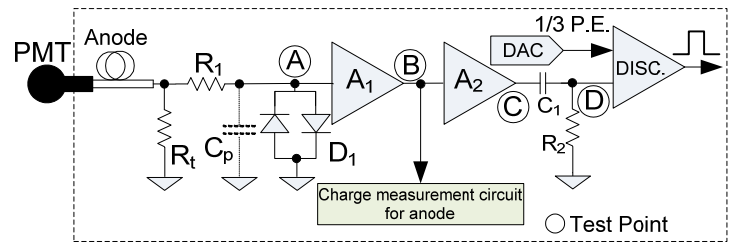

**Figure 3.** A leading edge discrimination circuit.

A circuit shown in figure Figure 3 has been designed, fabricated and tested. The test signals were generated by an arbitrary waveform generator (AFG3252, from Tektronix Inc.) so as to reproduce real PMT signals. Firstly, we evaluated the time measurement resolution performance based on the "cable delay method" [8]. The input signal was split and provided to two time discrimination circuits with the same circuit structure, and then the output pulses were measured by the FPGA-TDC (bin size: ~0.333 ns). Considering the situation where the time results of these two channels were not interrelated, the single-channel time resolution could be obtained by dividing by the RMS value of the time interval by $\sqrt{2}$. Test results indicate that the time resolution is better than 300 ps RMS over the whole dynamic range, which meets the application requirement with an excess.

During subsequent measurements the circuit dead time performance was evaluated carefully. Input of the circuit was stimulated with signal equivalent to around 4000 P.E. (voltage amplitude: ~5 V, rise time: ~4 ns, fall time: ~12 ns), and the resulting waveforms at the key circuit nodes are shown in figure Figure 4. All waveforms were acquired by an oscilloscope (LeCroy Inc., 104MXi). For the convenience of observation, collected data was processed by the MATLAB platform offline to normalize the signal baseline and arrival time to zero (the same below). Test results indicate that the bottom width ($T_b$) of the signal at the input of the discriminator widens to be more than 100 ns. A major reason is the high impedance of the protection diodes ($D_1$) discharge circuit to ground caused by $R_1$. In addition, a large signal under shoot and long recovery time are introduced by the AC coupled circuit, which also deteriorates the circuit dead time performance.

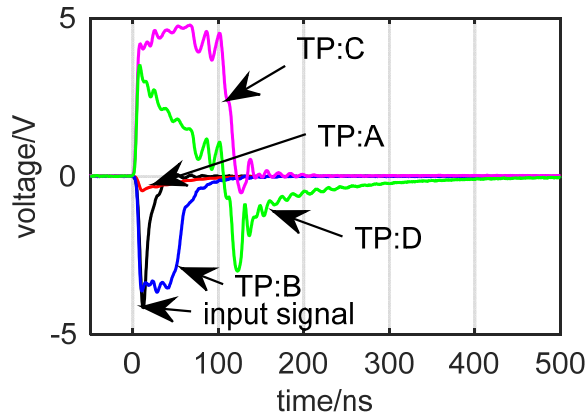

**Figure 4.** Waveforms at the key circuit nodes. Test points (TP: A, B, C, D) are referred to the corresponding circuit nodes in figure Figure **3**. (the same below)

Moreover, we evaluate the circuit dead time performance in the whole dynamic range. We use the ArbExpress software [9] to generate signal consisting of two pulses with an adjustable interval time ($T_{int}$). The first pulse shape was equivalent to 4000 P.E. and the second one was simulating 1 P.E. The $T_{int}$ interval was decreased until the latter pulse can't be discriminated. Figure Figure 5 presents limit case confirming that the dead time in the whole dynamic range is about 550 ns.

Further efforts have been made to optimize the circuit dead time, which will be discussed next.

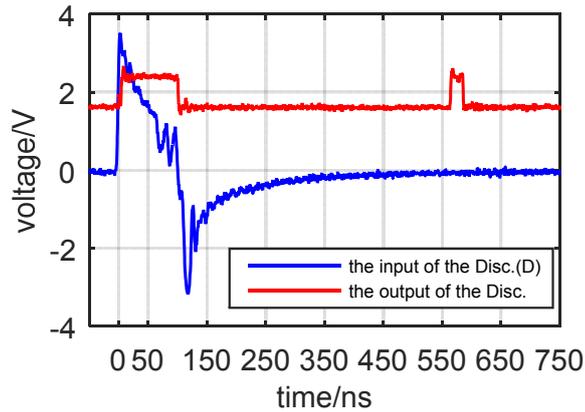

**Figure 5.** Dead time test results in the whole dynamic range.

**2.2 A new time discrimination circuit**

As the test results discussed above showed, there are two main factors that deteriorate the dead time performance of the previous circuit depicted in figure Figure 3:
(1) The high impedance of the protection diodes ($D_1$) discharge circuit to ground caused by $R_1$;
(2) A large signal under shoot and long recovery time at the input of the discriminator caused by the AC coupled circuit.

To solve the above two issues, a new discrimination circuit shown in figure Figure 6 was designed. Firstly, we design a diode discharge circuit to solve issue (1). The output of $A_2$ is split and connected to another discriminator (Disc.2, ADCMP601, from ADI Inc.). Output of this additional discriminator is used to control a RF switch (ADG918, from ADI Inc.). Once the RF switch is turned on, the protection circuit can discharge through a low resistance ($R_p$, 50 Ω) to ground. In addition, in order to control the turn-on time of the RF switch, a monostable circuit (74LVC1G123, from NXP Inc.) is used. Moreover, only the circuit input signals larger than 0.5 V shall be discriminated by Disc.2, otherwise the charge measurement circuit for anode is affected. The circuit dead time is determined mainly by the propagation delay from input to the RF switch ($T_{pd}$) and the turn-on time of the switch ($T_{turn-on}$). In order to make the circuit dead time as small as possible, $T_{pd}$ and $T_{turn-on}$ should be set small. However, $T_{pd}$ should not be too small to ensure that the leading edge of the input signal will not be corrupted. In our design, an analog delay line (LDH542N50BAA, from muRate Inc.) is used to adjust $T_{pd}$. According to the datasheets of the concerned chips, we can estimate that $T_{pd}$ is between around 12 ns and 30 ns (not including the analog delay line). Since the leading edge of the input signal is very fast, the propagation delay should be able to meet the design requirements. The monostable circuit can output pulse having minimal width of around 40 ns, so the theoretical circuit dead time should be less than 60 ns.

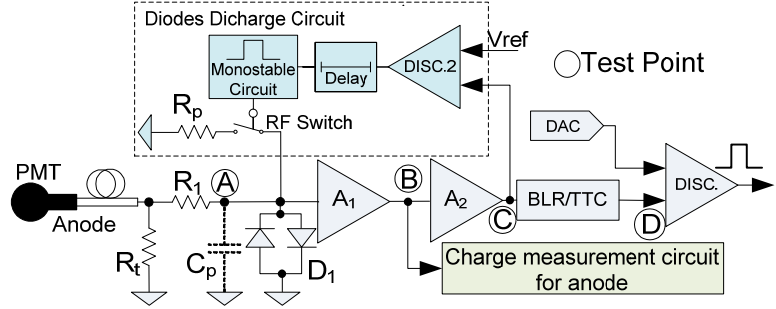

**Figure 6.** A new time discrimination circuit.

In order to verify the above analysis, Pspice simulation has been conducted. The transient waveform simulation results (input signal: ~4000 P.E.) are shown in figure Figure 7. The simulation results indicate that the discharge time of $D_1$ and the recovery time decrease noticeably, which all meet our design expectations.

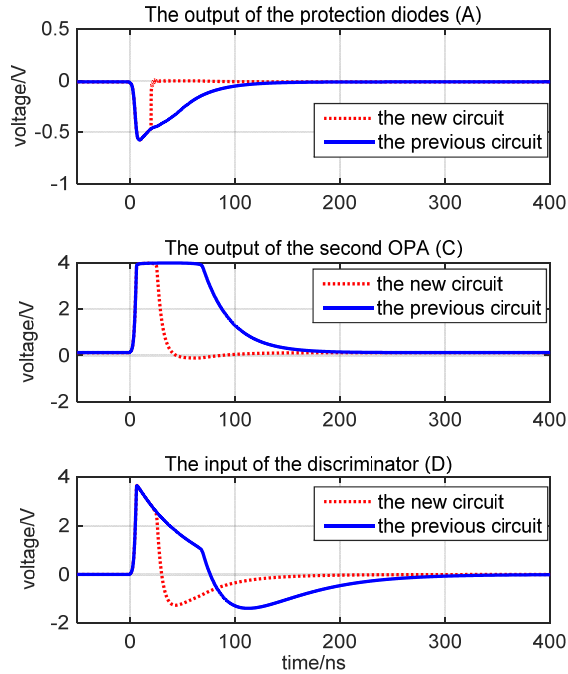

**Figure 7.** Comparison of the two designs based on transient simulation results.

As presented above, the AC coupled circuit would cause a large under shoot and long recovery time at the input of the discriminator. Therefore, a DC baseline restorer (BLR) is required, and many solutions have been studied. For instance, Cesare Liguori and Gianluigi Pessina presented a new self-buffered DC baseline restorer in Ref [10]; Claudio Arnaboldi and Gianluigi Pessina presented a new fast BLR in Ref [11], etc. These BLRs feature quasi-ideal behavior but have very complex circuit structure, and the use of the extra components increases circuit noise and decreases the circuit bandwidth, which would deteriorate the time resolution performance. Moreover, the offset voltage drift of the last-stage amplifier is hardly eliminated. Considering these reasons, in our design, a simple non-active CDR DC Baseline Restorer (BLR) [12] is employed at the input of the discriminator, as shown in figure Figure 8. For the under-shoot signals, the Schottky diode ($D_2$) turns on and the baseline recovers quickly. The Pspice simulation results indicate that the signal baseline at the input of the discriminator recovers more quickly due to the CDR BLR, as shown in figure Figure 9.

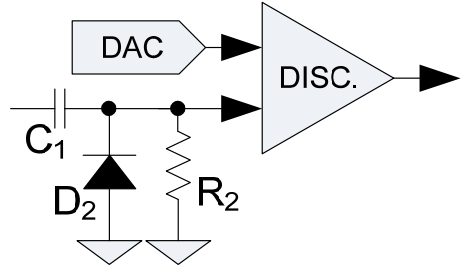

**Figure 8.** A simple non-active CDR DC Baseline Restorer.

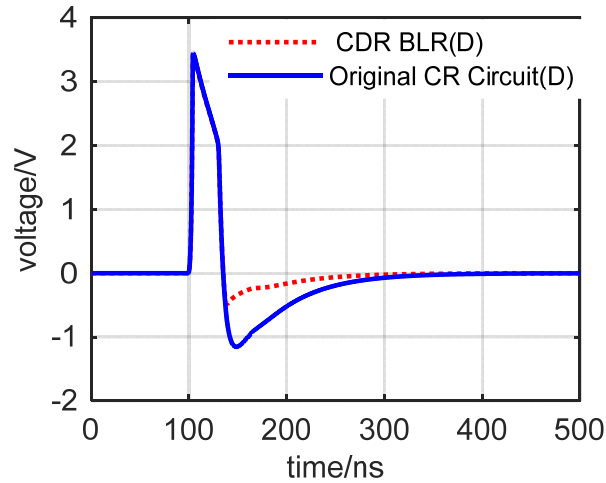

**Figure 9.** Transient waveform simulation results of the CDR BLR.

Besides the CDR BLR, attempt has been made to solve the issue (2) in another way. We wanted to remove the AC coupled circuit but for the ordinary leading edge discriminators, the threshold-level crossing of the signal at the input of the discriminator would walk with the drift of the signal baseline. Therefore, we designed a circuit named Threshold Tracking Circuit (TTC) which can adjust the threshold voltage automatically. The structure of TTC is shown in figure Figure 10. The output of $A_2$ is fed to a low-pass filter with large time constant ($R_4 \times C_2$) to acquire the voltage amplitude of the signal baseline ($V_B$), followed by a precise and zero-drift amplifier ($A_3$, ADA4528, from ADI Inc.) with a close-loop DC gain of $\beta$. Two high-precision resistors ($R_5$ and $R_6$) are employed to divide the voltage level between the amplified signal baseline ($\beta V_B$) and the output of DAC ($V_{DAC}$) to acquire the threshold voltage ($V_{th}$).

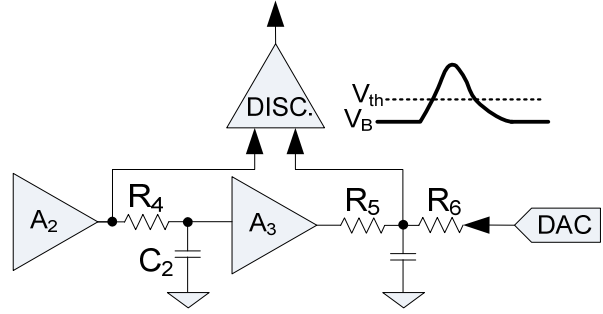

**Figure 10.** Structure of the Threshold Tracking Circuit.

Theatrically, the threshold voltage can be expressed as:

$$V_{th} = \frac{\beta R_6}{R_5 + R_6} V_B + \frac{R_5}{R_5 + R_6} V_{DAC} \quad , \qquad (1)$$

In order to avoid the baseline dependence of the threshold-level crossing of the signals, the following equation should be established:

$$\frac{\partial (V_{th} - V_B)}{\partial V_B} = 0 \quad , \qquad (2)$$

In our design, β is selected to be 2, and according to (1) and (2), $R_5$ and $R_6$ are set to the same value of 10 kΩ.

We conducted the Pspice simulation to verify the above analysis. We change the DC offset of the circuit input signal (1 P.E.) and the threshold voltage can adjust automatically to maintain the ($V_{th}$-$V_B$) fixed basically.

## 3. Performance Tests in the Lab

An improved time discrimination circuit shown in figure Figure 6 has been fabricated and tested. In the below Sub-sections, the performance of the diode discharge circuit, CDR and TTC will be evaluated in the lab separately.

### 3.1 Diode discharge circuit

Firstly, we tested the new time discrimination circuit with the diode discharge circuit and the amplified signals are simply AC coupled into the discriminator without the employment of the BLR or TTC. The waveforms at the key circuit nodes of the new circuit compared with those of the previous time discrimination circuit are shown in figure Figure 11. Both circuits were stimulated with input signal of around 4000 P.E. As can be easily seen, the diode discharge time in improved circuit decreases greatly. Pulse at the input of the discriminator $T_b$ shortens to around 50 ns and the recovery time decreases accordingly.

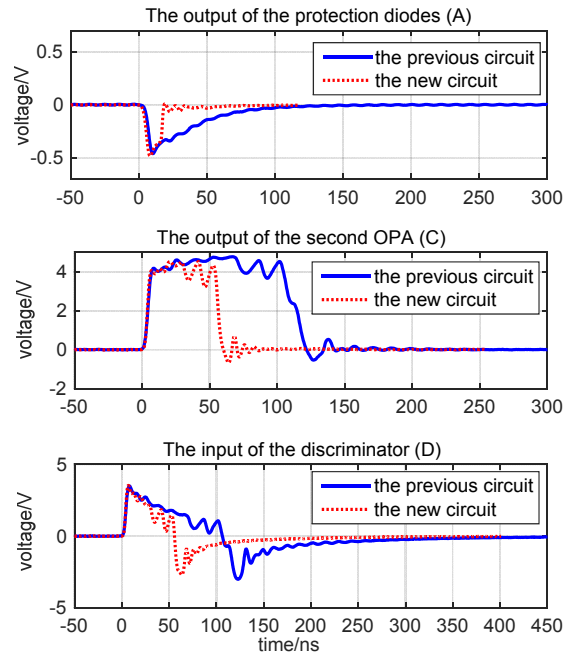

**Figure 11.** Transient waveforms test results of the new and previous circuits.

Figure Figure 12 shows the propagation delay from the circuit input to the input of the monostable circuit (around 25 ns) and the pulse width at the output of the monostable circuit (around 30 ns).

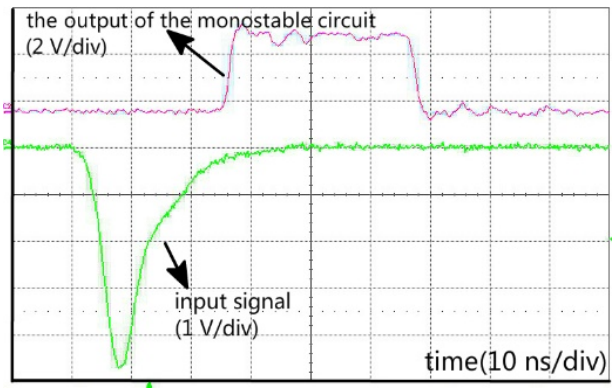

**Figure 12.** Oscilloscope screenshot of the circuit input signal and the output of the monostable circuit.

Besides, in order to evaluate the circuit dead time performance, input of the circuit was stimulated with two pulses with an adjustable interval time ($T_{int}$). The first pulse shape was equivalent to 4000 P.E., and we conducted the three tests as follows:
(A) $T_{int}$ is set to 60 ns, and width of second pulse is decreased until it is no longer discriminated;
(B) $T_{int}$ is set to 100 ns, and width of second pulse is decreased until it is no longer discriminated;
(C) Second pulse is set to be equivalent to 1 P.E. $T_{int}$ is decreased as long as the second pulse is correctly discriminated. Obtained value is a dead time in the whole dynamic range.

The test results are shown in figure Figure 13. Test results indicate that the minimum discriminable signal is about 60 P.E. for the interval of 60 ns while 20 P.E. for the interval of 100 ns. The dead time in the whole dynamic range is about 300 ns. Comparing with the previous circuit without the diode discharge circuit, the circuit dead time performance increases.

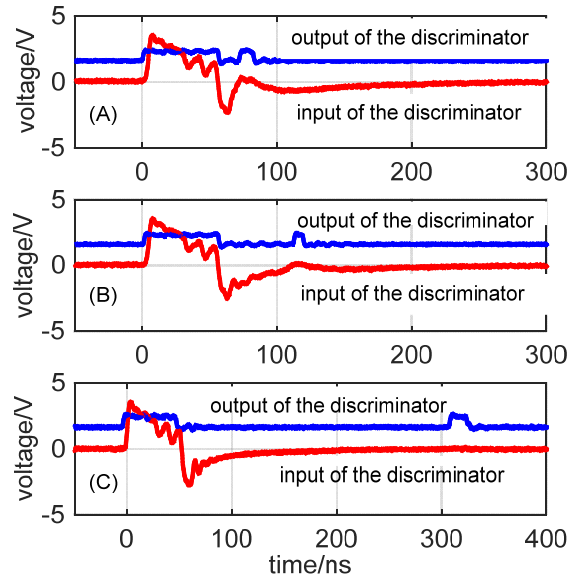

**Figure 13.** Dead time test results of the new circuit (diode discharge circuit).

### 3.2 CDR BLR

In the tests described above, the signals were directly AC coupled into the discriminator. Following is the discussion of measurements results taken with the CDR BLR mentioned in section 2 combined with the diode discharge circuit. Figure Figure 14 shows the waveforms at the input of the discriminator. As expected, the use of the CDR BLR circuit makes the signal baseline recover much faster than in the simple AC coupled circuit.

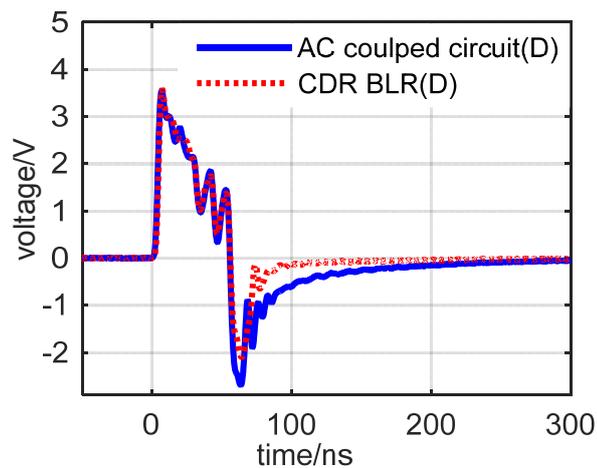

**Figure 14.** Waveforms at the input of the discriminator.

We also executed the three dead time tests (A, B, C) mentioned in section 3.1 and the results are depicted in figure Figure 15. Test results indicate that the minimum discriminable signal is about 40 P.E. for the interval of 60 ns while 5 P.E. for the interval of 100 ns. The dead time in the whole dynamic range is about 200 ns. Therefore, the CDR BLR effectively improves the circuit dead time performance.

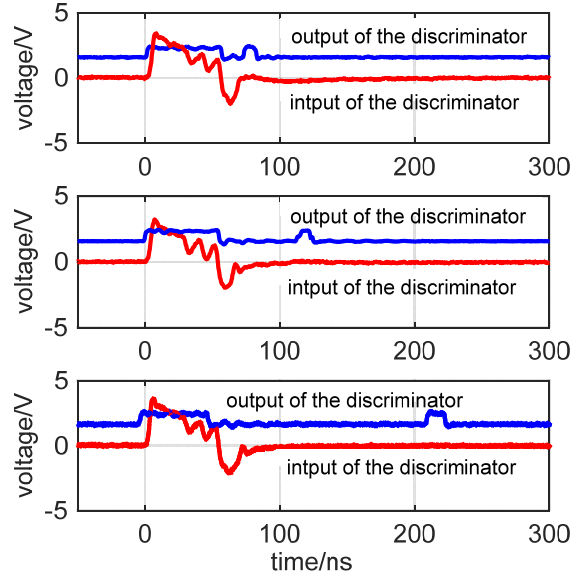

**Figure 15.** Dead time test results of the new circuit (diode discharge circuit with CDR BLR).

### 3.3 TTC

We also tested the TTC presented in section 2 combined with the diode discharge circuit. We changed the DC offset of the input signal and the threshold voltage could auto-adjust to guarantee the stability of ($V_{th}$-$V_B$). However, it addresses a serious problem we have found that the input offset voltage ($V_{in-os}$) at the input of the amplifier ($A_1$, marked in figure Figure 6) is very large (~ +10 mV). Even if the RF switch pulls the voltage level to ground quickly, the signal baseline would still recover to $V_{in-os}$, which would introduce a slow recovery signal after the fall edge at the input of the discriminator, shown in figure Figure 16.

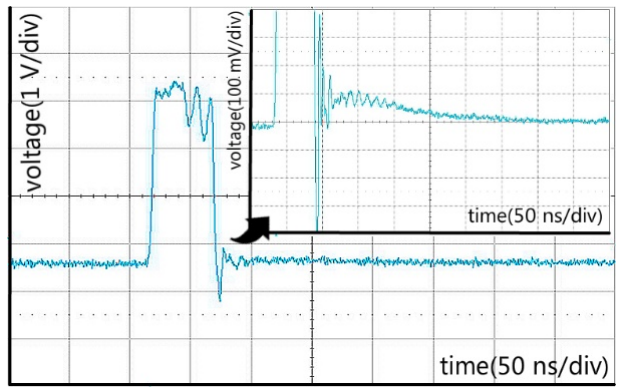

**Figure 16.** Oscilloscope screenshot of signal at the input of the discriminator (TTC).

The output of the discriminator shown in figure Figure 17 is plotted in the afterglow mode of the oscilloscope screen. It shows that a long dead time is introduced after the RF switch is turned on. Therefore, the TTC has no advantage for the dead time performance improvement in this enhanced time measurement circuit.

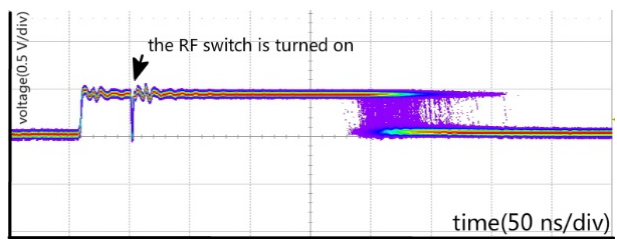

**Figure 17.** Oscilloscope screenshot of the output of the discriminator (afterglow mode, TTC).

**3.4 Time resolution**

According to the above analysis and test results, the diode discharge circuit combined with the CDR BLR is finally employed to enhance the circuit dead time performance in the new time discrimination circuit. Furtherly, we also compared the time resolution performance of the new time discrimination circuit and the previous circuit version. Results are shown in figure Figure 18.

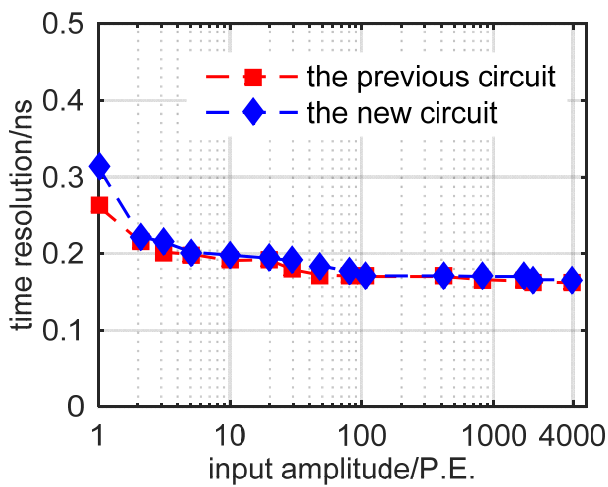

**Figure 18.** Time resolution test results of the new and previous circuits.

According to figure Figure 18, the time resolution @ 1 P.E. of the new time discrimination circuit deteriorates from 270 ps RMS to 310 ps RMS while for higher number of P.E. the time resolution is almost the same. Nevertheless, obtained result meets the application requirements. The main reason of the deterioration is that the circuit front-end parasitic capacitance increases due to the introduction of the RF switch, which slows the signal edge rate.

**3.5 Multi-channel test**

So far we have tested 6 channels with the improved circuit structure of diode discharge circuit combined with the CDR BLR. Each channel has a dead time less than 200 ns over the whole dynamic range, shown in figure Figure 19. Furthermore, for each channel, $T_{pd}$ is less than 30 ns and $T_{turn-on}$ is around 30 ns, which all meet our design expectations. We also evaluated the time resolution performance. The time resolution of each channel agrees with the curve shown in figure Figure 18, which is beyond our application requirements.

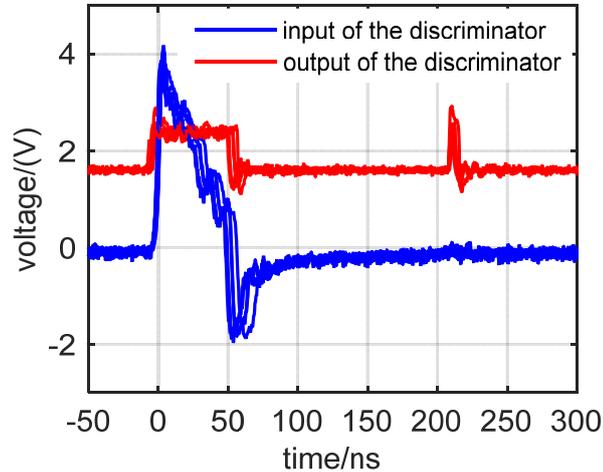

**Figure 19.** multi-channel dead time test results.

## 4. Conclusion

In this paper, key techniques in a new time discrimination circuits designed for LHAASO WCDA were discussed. We propose several approaches to optimize the circuit dead time performance. Through circuit analysis, simulations and tests, circuit optimization was made, and a good time resolution and a small dead time of around 200 ns was achieved in a large dynamic range of 1 P.E. to 4000 P.E..

**Acknowledgments**

This work is supported by Knowledge Innovation Program of the Chinese Academy of Sciences (KJCX2-YW-N27), National Natural Science Foundation of China (11175174) and the CAS Center for Excellence in Particle Physics (CCEPP).

**References**

[1] C. Zhen, *A future project at tibet: the large high altitude air shower observatory (LHAASO)*, Chin.Phys.C, 34(2010)249
[2] H. H. He et al. *LHAASO Project: Detector Design and Prototype*, in *Proceedings of the 31st ICRC, 2009, LODZ*, Poland.
[3] M.J. Chen et al, *R&D of LHAASO-WCDA, in procedings of 32nd International Cosmic Ray Conference,* 2011, Beijing China.
[4] L. Zhao et al, *Proposal of the readout electronics for the WCDA in the LHAASO experiment,* Chin.Phys.C, 38(2014)016101
[5] C. Ma, L. Zhao et al. *Analog front-end prototype electronics for the LHAASO WCDA*, Chin. Phys. C, 40(2016): 016101.
[6] C.F. Ye, L. Zhao et al, *A field-programmable-gate-array based time digitizer for the time-of-flight mass Spectrometry, Rev. Sci. Instrum., 85(2014) 045115*

[7]   http://www.analog.com/media/en/technical-documentation/data/sheets/AD8000.pdf, retrieved 2[th] August 2016

[8] L.F. Kang, L. Zhao et al, *A 128-channel high precision time measurement module*, Metrol. Meas. Syst., *XX(2013)275*.

[9] http://cn.tek.com/product-software series/ arbexpress-signal-generator-software, retrieved 2[th] August 2016

[10] C. Liguori, G. Pessina, *A self-buffered DC baseline restorer with quasi-ideal behavior, Nucl. Instrum. and Meth. in Physics Research A 437(1999)557.*

[11] C. Arnaboldi and Gianluigi, *A very simple baseline restorer for nuclear applications, Nucl. Instrum. and Meth. in Physics Research A 512(2003)129.*

[12] A. F. Arbel, *Analog signal processing and instrumentation*,: Cambridge university press, London 1984.